\def\rn{}
\def\nn#1 #2{#2. #1}				
\def\nnn#1 #2 #3{#2. #3. #1}			
\def\nnnn#1 #2 #3 #4{#2. #3. #4 #1}		
\def\nnnnn#1 #2 #3 #4 #5{#2. #3. #4 #5. #1}	

\def\rf#1;#2;#3;#4;#5 {{\frenchspacing\par\rn#1, #3 {\bf #4}, #5 (#2). \par}}
\def\rg#1;#2;#3;#4;#5;#6 {{\frenchspacing\par\rn#1, #3 {\bf #4}, #5 (#2). \par}}
\def\rfbook#1;#2;#3;#4;#5 {{\frenchspacing\par\rn#1, {\it #3} (#5, #4, #2).\par}}
\def\rfprep#1;#2;#3 {{\par\frenchspacing\rn#1, #3 (#2).\par}}
\def\rfproc#1;#2;#3;#4;#5;#6 {{\frenchspacing\par\rn#1 #2, in {\it #3}, ed. #4 (#5: #6)\par}}
\def\rfprocp#1;#2;#3;#4;#5;#6;#7 {{\frenchspacing\par\rn#1 #2, in {\it #3}, ed. #4 (#5: #6), p#7\par}}

\def\rg#1;#2;#3;#4;#5;#6 {\par\rn#1 #2, {\it #3}, {\bf #4}, #5 (``#6'') \par}
\def\rf#1;#2;#3;#4;#5 {\par\rn#1, {\it #3}, {\bf #4}, #5 (#2)\par}
\def\rfbook#1;#2;#3;#4;#5 {{\frenchspacing\par\rn#1, {\it #3} (#4: #5, #2)\par}}
\def\rfproc#1;#2;#3;#4;#5;#6 {{\frenchspacing\par\rn#1 #2, in {\it #3}, ed. #4 (#5: #6)\par}}
\def\rfprocp#1;#2;#3;#4;#5;#6;#7 {{\frenchspacing\par\rn#1 #2, in {\it #3}, ed. #4 (#5: #6), p#7\par}}
\def\rfprep#1;#2;#3  {{\par\rn#1, #3 (#2)\par}}
\def\rfprepp#1;#2;#3 {{\par\rn#1 #2, #3\par}}

\def\fig#1{Figure~\ref{#1}}

\documentclass[twocolumn,amsmath,nofootinbib]{revtex4} 
\usepackage{url}
\begin{document}
\input{epsf.sty}

\title{Many lives in many worlds}
\author{Max Tegmark}
\address{(In this universe:) Dept.~of Physics \& MIT Kavli Institute, Massachusetts Institute of Technology, Cambridge, MA 02139, USA}
\date{Published in {\it Nature}, {\bf 448}, 23, July 2007}

\begin{abstract}
I argue that accepting quantum mechanics to be universally true means that you should
also believe in parallel universes.
I give my assessment of Everett's theory as it celebrates its 50th anniversary.
\end{abstract}
  
\maketitle


\begin{figure}[pbt]
\centerline{{\vbox{\epsfxsize=8.7cm\epsfbox{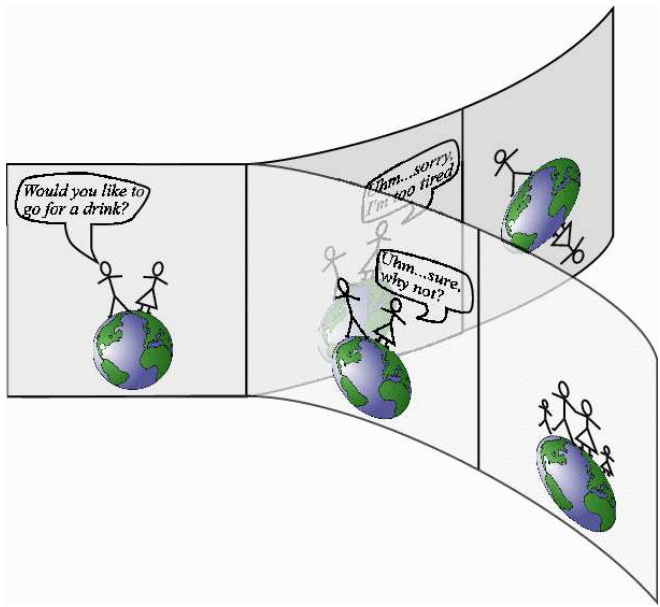}}}}
\smallskip
\caption{
Is it only in fiction that we can experience parallel lives? If atoms can be in two places at once, so can you.}
\label{CartoonFig}
\end{figure}
Almost all of my colleagues have an opinion about it, but almost none of them 
have read it. The first draft of Hugh 
Everett's PhD thesis, the shortened official 
version of which celebrates its 50th birthday 
this year, is buried in the out-of-print book 
{\it The Many-Worlds Interpretation of Quantum 
Mechanics} \cite{EverettBook}. I remember my excitement on 
finding it in a small Berkeley book store back 
in grad school, and still view it as one of the most brilliant texts I've ever read. 

By the time Everett started his graduate work with John Archibald Wheeler at Princeton 
University, quantum mechanics had chalked up stunning successes in explaining the 
atomic realm, yet a debate raged on as to what its mathematical formalism really meant. I 
was fortunate to get to discuss quantum mechanics with Wheeler during my postdoctorate 
years in Princeton, but never had the chance to meet Everett.

Quantum mechanics specifies the state of the universe not in classical terms, such as the 
positions and velocities of all particles, but in terms of a mathematical object called a 
wavefunction. According to the Schr{\"o}dinger equation, this wavefunction evolves over 
time in a deterministic fashion that mathematicians term ``unitary". Although quantum 
mechanics is often described as inherently random and uncertain, there is nothing random 
or uncertain about the way the wavefunction evolves.

The sticky part is how to connect this wavefunction with what we observe. Many 
legitimate wavefunctions correspond to counterintuitive situations, such as Schr{\"o}dinger's 
cat being dead-and-alive at the same time in a ``superposition" of states. In the 1920s, 
physicists explained away this weirdness by postulating that the wavefunction 
``collapsed" into some random but definite classical outcome whenever someone made an 
observation. This add-on had the virtue of explaining observations, but rendered the 
theory incomplete, because there was no mathematics specifying what constituted an 
observation -- that is, when the wavefunction was supposed to collapse.

\begin{figure}[tbp]
\centerline{{\vbox{\epsfxsize=8.6cm\epsfbox{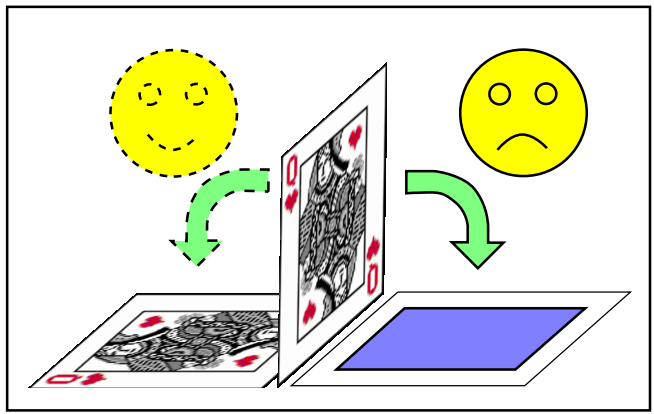}}}}
\bigskip
\centerline{{\vbox{\epsfxsize=8.6cm\epsfbox{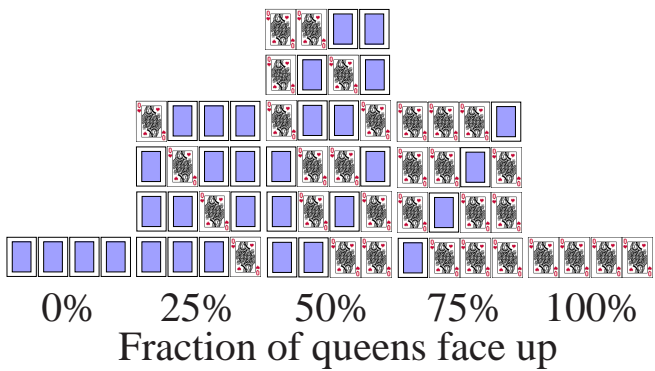}}}}
\bigskip
\caption{
According to quantum theory, a card perfectly balanced on its edge will fall down in what 
is known as a ``superposition" --- the card really is in two places at once. If a gambler bets 
money on the queen landing face up, her own state changes to become a superposition of 
two possible outcomes --- winning or losing the bet. In either of these parallel worlds, the 
gambler is unaware of the other outcome and feels as if the card fell randomly.
If the gambler repeats this game four times in a row, there will be 16 ($2\times 2\times 2\times 2$) possible 
outcomes, or parallel worlds. In most of these worlds, it will seem that queens occur 
randomly, with about 50\% probability. Only in two worlds will all four cards land the 
same way up. If the game is continued many more times, almost every gambler in each of 
the worlds will conclude that the laws of probability apply even though the underlying 
physics is not random and, as Einstein would have put it, ``God does not play dice".
}
\label{CardFig}
\end{figure}

Everett's theory is simple to state but has complicated implications, including parallel 
universes.  The theory can be summed up by saying that the Schr{\"o}dinger equation applies 
at all times; in other words, that the wavefunction never collapses. That's 
it -- no mention of parallel universes or splitting worlds, which are implications of the 
theory rather than postulates.  His brilliant insight was that this collapse-free quantum 
theory is, in fact, consistent with observation. Although it predicts that a wavefunction 
describing one classical reality gradually evolves into a wavefunction describing a 
superposition of many such realities -- the many worlds -- observers subjectively 
experience this splitting merely as a slight randomness (see \fig{CardFig}), with probabilities 
consistent with those calculated using the wavefunction-collapse recipe.

\bigskip{\bf Gaining acceptance}

It is often said that important scientific 
discoveries go though three phases: first they are completely ignored, then they are 
violently attacked, and finally they are brushed aside as well-known. Everett's discovery 
was no exception: it took over a decade until it started getting noticed. 
But it was too late for Everett who left academia disillusioned \cite{bio}.

Everett's no-collapse idea is not yet at stage three, but after being widely dismissed as too 
crazy during the 1970s and 1980s, it has gradually gained more acceptance. At an 
informal poll taken at a conference on the foundations of quantum theory in 1999 
physicists rated the idea more highly than the alternatives, although there were still many 
more "undecided" \cite{quantum}. I believe the upwards trend is clear.

Why the change? I think there are several reasons. Predictions of other types of parallel 
universes from cosmological inflation and string theory have increased tolerance for 
weird-sounding ideas. New experiments have demonstrated quantum weirdness in ever 
larger systems. Finally, the discovery of a process known as decoherence has answered 
crucial questions that Everett's work had left dangling.

For example, if these parallel universes exist, then why don't we perceive them? Quantum 
superpositions cannot be confined -- as most quantum experiments are -- to the 
microworld. Because you are made of atoms, then if atoms can be in two places at once 
in superposition, so can you (\fig{CartoonFig}).

The breakthrough came in 1970 with a seminal paper by H.~Dieter Zeh, who showed that the 
Schr{\"o}dinger equation itself gives rise to a type of censorship. This effect became known 
as ``decoherence", and was worked out in great detail by Wojciech Zurek, Zeh and others 
over the following decades. Quantum superpositions were found to remain observable 
only as long as they were kept secret from the rest of the world. The quantum card in  
\fig{CardFig} is constantly bumping into air molecules, photons, and so on, which 
thereby find out whether it has fallen to the left or to the right, destroying the coherence 
of the superposition and making it unobservable. Decoherence also explains why states 
resembling classical physics have special status: they are the most robust to decoherence.

\bigskip{\bf Science of philosophy?}

The main motivation for introducing the notion of random wavefunction collapse into 
quantum physics had been to explain why we perceive probabilities and not strange 
macroscopic superpositions. After Everett had shown that things would appear random 
anyway (\fig{CardFig}) and decoherence had been found to explain why we never perceived 
anything strange, much of this motivation was gone. Even though the wavefunction 
technically never collapses in the Everett view, it is generally agreed that decoherence 
produces an effect that looks like a collapse and smells like a collapse.
 
In my opinion, it is time to update the many quantum textbooks that introduce 
wavefunction collapse as a fundamental postulate of quantum mechanics. The notion of 
collapse still has utility as a calculational recipe, but students should be told that it is 
probably not a 
fundamental process violating the Schr{\"o}dinger equation so as to avoid any subsequent 
confusion. If you are considering a quantum textbook that does not mention ``Everett" and 
``decoherence" in the index, I recommend buying a more modern one.

After 50 years we can celebrate the fact that Everett's interpretation is still consistent with 
quantum observations, but we face another pressing question: is it science or mere 
philosophy? The key point is that parallel universes are not a theory in themselves, but a 
prediction of certain theories. For a theory to be falsifiable, we need not observe and test 
all its predictions -- one will do.

Because Einstein's theory of General Relativity has successfully predicted many things 
that we can observe, we also take seriously its predictions for things we cannot, like the 
internal structure of black holes. Analogously, successful predictions by unitary quantum 
mechanics have made scientists take more seriously its other predictions, including 
parallel universes

Moreover, Everett's theory is falsifiable by future lab experiments: no matter how large a 
system they probe, it says, they will not observe the wavefunction collapsing. Indeed, 
collapse-free superpositions have been demonstrated in, for example, carbon-60 
molecules. Several groups are now attempting to create quantum superpositions of 
objects involving $10^17$ atoms or more, tantalizingly close to our human macroscopic 
scale. There is also a global effort to build quantum computers which, if successful, will 
be able to factor numbers exponentially faster than classical computers, effectively 
performing parallel computations in Everett's parallel worlds.

\bigskip{\bf The bird perspective}

So Everett's theory is testable and so far agrees with observation. But should you really 
believe it?  When thinking about the ultimate nature of reality, I find it useful to 
distinguish between two ways of viewing a physical theory: the outside view of a 
physicist studying its mathematical equations, like a bird surveying a landscape from 
high above, and the inside view of an observer living in the world described by the 
equations, like a frog being watched by the bird.

From the bird perspective, Everett's multiverse is simple. There is only one wavefunction, 
and it evolves smoothly and deterministically over time without any kind of splitting or 
parallelism. The abstract quantum world described by this evolving wavefunction 
contains within it a vast number of classical parallel storylines (``worlds"), continuously 
splitting and merging, as well as a number of quantum phenomena that lack a classical 
description. From their frog perspective, observers perceive only a tiny fraction of this 
full reality, and they perceive the splitting of classical storylines as quantum randomness. 

What is more fundamental -- the frog perspective or the bird perspective? In other words, 
what is more basic to you: human language or mathematical language? If you opt for the 
former, you would probably prefer a ``many words" interpretation of quantum mechanics, 
where mathematical simplicity is sacrificed to collapse the wavefunction and eliminate 
parallel universes.

But if you prefer a simple and purely mathematical theory, then you -- like me -- are stuck 
with the many worlds interpretation. If you struggle with this you are in good company: 
in general, it has proven extremely difficult to formulate a mathematical theory that 
predicts everything we can observe and nothing else -- and not just for quantum physics.

Moreover, we should expect quantum mechanics to feel counterintuitive because 
evolution endowed us with intuition only for those aspects of physics that had survival 
value for our distant ancestors, such as the trajectories of flying rocks.

The choice is yours. But I worry that if we dismiss theories like Everett's because we 
can't observe everything or because they seem weird, we risk missing true breakthroughs, 
perpetuating our instinctive reluctance to expand our horizons. To modern ears the 
Shapley-Curtis debate of 1920 about whether there were really a multitude of galaxies 
(parallel universes by the standards of the time) sounds positively quaint. 

Everett asked us to acknowledge that our physical world is grander than we had 
imagined, a humble suggestion that is probably easier to accept after the recent 
breakthroughs in cosmology than it was 50 years ago. I think Everett's only mistake was 
to be born ahead of his time. In another 50 years, I believe we will be more used to the 
weird ways of our cosmos, and even find its strangeness to be part of its charm.

\bigskip
{\bf Acknowledgments:}
This work was supported by NASA grant and NNG06GC55G,
NSF grants AST-0134999 and 0607597, the Kavli Foundation, and fellowships from the David and Lucile
Packard Foundation and the Research Corporation.


\end{document}